\documentstyle [epsf,spie,twocolumn] {article}

\title {Chandra X-Ray Observatory (CXO):  Overview\\ }

\author {M. C. Weisskopf\supit{a}, H. D. Tananbaum\supit{b}, L. P. Van Speybroeck\supit{b}, and S. L. O'Dell\supit{a},  
\skiplinehalf 
\supit{a} NASA Marshall Space Flight Center \skipline 
Huntsville, AL \hspace{0.5em}35812 \hspace{0.5em}USA 
\skiplinehalf 
\supit{b} Smithsoniam Astrophysical Observatory \skipline
Cambridge, MA 02138 USA 
}

\authorinfo{{Invited Paper, in {\em X-Ray Optics, Instruments, and Missions},  J.Truemper and B. Aschenbach, eds., {\em Proc.\ SPIE}\/ {\bf 4012}, 2000.}\\
M.C.W.: martin@smoker.msfc.nasa.gov; 256-544-7740. \\
H.D.T.: ht@cfa.harvard.edu; 617-495-7248. \\
L.V.S.: lvs@cfa.harvard.edu; 617-495-7233. \\
S.L.O.: odell@cosmos.msfc.nasa.gov; 256-544-7708. }

\newcommand{\eg}{{\em e.g.}, }

\newcommand{\lae}{\mathrel{<\kern-1.0em\lower0.9ex\hbox{$\sim$}}}
\newcommand{\gae}{\mathrel{>\kern-1.0em\lower0.9ex\hbox{$\sim$}}}

\begin{document}
\maketitle

\begin {abstract}
The Chandra X-Ray Observatory (CXO), the x-ray component of NASA's Great Observatories, was launched early in the morning of 1999, July 23 by the Space Shuttle {\sl Columbia}.  
The Shuttle launch was only the first step in placing the observatory in orbit.  
After release from the cargo bay, the Inertial Upper Stage performed two firings, and separated from the observatory as planned.  
Finally, after five firings of Chandra's own Integral Propulsion System~--- the last of which took place 15 days after launch~--- the observatory was placed in 
its highly elliptical orbit of $\sim$140,000 km apogee and $\sim$10,000 km perigee.  
After activation, the first x-rays focussed by the telescope were observed on 1999, August 12.  
Beginning with these initial observations one could see that the telescope had survived the launch environment and was operating as expected.  
The month following the opening of the sunshade door was spent adjusting the focus for each set of instrument configurations, determining the optical axis, calibrating the star camera, establishing the relative response functions, determining energy scales, and taking a series of "publicity" images.  
Each observation proved to be far more revealing than was expected.  
Finally, and despite an initial surprise and setback due to the discovery that the Chandra x-ray telescope was far more efficient for concentrating low-energy protons than had been anticipated, the observatory is performing well and is returning superb scientific data.  
Together with other space observatories, most notably the recently activated XMM-Newton, it is clear that we are entering a new era of discovery in high-energy astrophysics.


\end {abstract}

\section {Introduction} \label{s:introduction}

The Chandra X-Ray Observatory (CXO), formerly known as the Advanced X-Ray Astrophysics Facility (AXAF), has joined the Hubble Space Telescope (HST) and the Compton Gamma-Ray Observatory (CGRO) as one of NASA's "Great Observatories".  
CXO provides unprecedented capabilities for sub-arcsecond imaging, spectrometric imaging, and for high-resolution dispersive spectroscopy over the x-ray band 0.08-10 keV (15-0.12 nm).  
With these capabilites a wide variety of high-energy phenomena in a broad range of astronomical objects is being observed.

Chandra is a NASA facility which provides scientific data to the international astronomical community in response to scientific proposals for its use.  
The CXO is the product of the efforts of many commercial, academic, and government organizations in the United States and Europe.  
NASA Marshall Space Flight Center (MSFC, Huntsville, Alabama) manages the Project and provides Project Science; TRW Space and Electronics Group (Redondo Beach, California) served as prime contractor; the Smithsonian Astrophysical Observatory (SAO, Cambridge, Massachusetts) provides the technical support and is responsible for ground operations including the Chandra X-ray Center (CXC).  
There are also five scientific instruments aboard the Observatory provided by a number of different institutions.  
These instruments are discussed in \S\ref{ss:ephin} and \S\ref{ss:xraysys}.
\vspace{0.10in}

In 1977, NASA/MSFC and the Smithsonian Astrophysical Observatory (SAO) began the phase-A study leading to the definition of the then AXAF mission.
This study, in turn, had been intiated as a result of an unsolicited proposal submitted to NASA in 1976 by Prof. R. Giacconi and Dr. H. Tananbaum.
During the intervening years, several significant milestones transpired, including the highest recommendation by the National Academy of Sciences, selection of the instruments, selection of the prime contractor, demonstration of the optics, restructuring of the mission, and ultimately the launch (Figure~\ref{fig:launch})
\vspace{0.10in}

\begin {figure} [htb]

\caption {\label{fig:launch} 
Photograph of the {\sl Columbia} launch with the CXO payload (STS-93).  NASA Photo.}
\end {figure}

We begin by describing the Chandra systems (\S\ref{s:systems}) and the ground calibration (\S\ref{s:calibration}).
We then describe Chandra's on-orbit performance and its ability (\S\ref{s:onorbit}) to serve as NASA's premier facility for x-ray astrophysics.

\section {Chandra Systems} \label{s:systems}

\subsection {Mission and Orbit} \label{ss:mission}

The Space Shuttle {\sl Columbia} carried and deployed the Chandra into a low earth orbit, as NASA's Space Transportation System mission STS-93.
About 8 hours after launch Chandra was deployed (Figure~\ref{fig:deploy}) at an altitude of about 240 km (130 nautical miles).  
At this time, an Inertial Upper Stage (IUS), a two-stage solid-fuel rocket booster developed by the Boeing Company Defense and Space Group (Seattle, Washington) for the US Air Force, propelled the Chandra flight system into a highly elliptical transfer orbit.
Subsequently, over a period of days, Chandra's Internal Propulsion System (IPS), built by TRW, placed the observatory into its initial operation orbit - 140-Mm (87,000-nautical-mile) apogee and 10-Mm (6,200-nautical-mile) perigee, with a 28.5$^o$ initial inclination.

Chandra's highly elliptical orbit, with a period of 63.5 hours, yields a high observing efficiency.
The fraction of the sky occulted by the earth is small over most of the orbital period, as is the fraction of the time when the detector backgrounds are high as the Observatory dips into Earth's radiation belts.
Consequently, more than 70\% of the time is useful and uninterrupted observations lasting more than 2 days are possible.

The specified design life of the mission is 5 years; however, the only expendable (gas for maneuvering) is sized to allow operation for more than 10 years.
The orbit will be stable for decades.
\vspace{0.10in}

\subsection {Flight system} \label{ss:flightsys}
\vspace{0.10in}

\begin {figure} [htb]

\caption {\label{fig:deploy} 
Chandra attached to the IUS (to the right) after deployment from {\sl Columbia} but before IUS separation.
Note that the apparent narrowing of the CXO optical bench is an illusion due to shadowing.  NASA Photo.}
\end {figure}

With the IUS attached (Figure~\ref{fig:deploy}), the Chandra was the largest and heaviest payload ever deployed from an STS space shuttle.
Once deployed and separated from the IUS, the Chandra flight system is 13.8-m (43.5-ft) long by 4.2-m (14-ft) diameter, with a 19.5-m (64-ft) solar-panel wingspan.
With extensive use of graphite-epoxy structures, the mass of the Chandra flight system is $\sim$4,800 kg (10,600 pounds).
The Chandra flight system (Figure~\ref{fig:flightsys}) itself has 3 major modules or systems~--- the Spacecraft Module (\S\ref{sss:scmodule}), the Telescope System (\S\ref{sss:hrmasys}), and the Integrated Science Instrument Module (\S\ref{sss:sisys}).

\begin {figure} [htb]

\caption {\label{fig:flightsys} 
Expanded view of the Chandra flight system, showing several subsystems of the 3 major modules~--- the Telescope System, the Integrated Science Instrument Module, and the Spacecraft Module.  TRW drawing.}
\end {figure}

\subsubsection {Spacecraft module} \label{sss:scmodule}

TRW Space and Electronics Group (Redondo Beach, Califronia) built the Spacecraft Module which is made up of:

\begin{itemize}
\item [1.]  The Pointing Control and Aspect Determination (PCAD) subsystem which performs on-board attitude determination, solar-array control, slewing, pointing and dithering control, and momentum management.
\item [2.]  The Communication, Command, and Data Management (CCDM) subsystem which performs communications, command storage and processing, data acquisition and storage, and computation support, timing reference, and switching of primary electrical power for other Chandra systems or subsystems.
\item [3.]  The Electrical Power Subsystem (EPS) which generates, regulates, stores, distributes, conditions, and controls the primary electrical power.
\item [4.]  The Thermal Control Subsystem (TCS) which furnishes passive thermal control (where possible), heaters, and thermostats.
\item [5.]  The structures and mechanical subsystem which encompasses the spacecraft structures, mechanical interfaces among the spacecraft subsystems and with the telescope system and external structures.
\item [6.]  The propulsion subsystem which comprises the Integral Propulsion Subsystem (IPS) - deliberately disabled once final orbit was obtained - and the Momentum Unloading Propulsion Subsystem (MUPS).
\item [7.]  The flight software which implements algorithms for attitude determination and control, command and telemetry processing and storage, and thermal and electrical power monitoring and control.
\end{itemize}

\subsubsection {Telescope system} \label{sss:hrmasys}

The Eastman Kodak Company (Kodak, Rochester, New York) integrated the Telescope System. 
Its principal subsytems are the High-Resolution Mirror Assembly (HRMA, \S\ref{sss:hrma}) and the Optical Bench Assembly (OBA).
Composite Optics Incorporated (COI, San Diego, California) developed the critical light-weight composite materials for the OBA (and for other Chandra structures).
The Telescope System also provides mounts and mechanisms for the Chandra Observatory's 2 objective transmission gratings (\S\ref{sss:gratings}).
In addition, Ball Aerospace and Technologies Corporation (Boulder, Colorado) fabricated the Aspect Camera Assembly~\cite{Michaels1998}, a visible-light telescope and CCD camera which attaches to, and is coupled with, the Telescope System through a fiducial-light Transfer System, which effectively maps the x-ray focal plane onto the sky.

\subsubsection {Integrated Science Instrument Module} \label{sss:sisys}

Ball Aerospace and Technologies Corporation (Boulder, Colorado) built the Science Instrument Module~\cite{Skinner1997} (SIM), which includes mechanisms for focussing  and translating Chandra's focal-plane science instruments (\S\ref{sss:fpsi}).
The translation is necessary as the instruments cannot realistically share the focal plane and must be translated into position at the HRMA focus.
The Integrated Science Instrument Module (ISIM) simply denotes the SIM with the 2 focal-plane science instruments integrated.

\subsection {Electron Proton Helium Intrument (EPHIN)} \label{ss:ephin}

Mounted on the spacecraft and near the HRMA is a particle detector called the Electron, Proton, Helium INstrument (EPHIN).
The EPHIN instrument was built by the Instit\"ut f\"ur Experimentelle und Angewandte Physik und Extraterrestrishce Physik at the University of Kiel, Germany.
The EPHIN detector is used to monitor the local charged particle environment as part of the scheme to protect the focal-plane instruments from particle radiation damage.
EPHIN consists of an array of 5 silicon detectors with anti-coincidence.
The instrument is sensitive to electrons in the energy range 150 keV - 5 MeV, and protons/helium isotopes in the energy range 5 - 49 MeV/nucleon.  The field of view is 83 degrees.
The forerunner of the Chandra-EPHIN was flown on the SOHO satellite.

\subsection {X-Ray Subsystems} \label{ss:xraysys}

Chandra's x-ray subsytems are the High-Resolution Mirror Assembly (HRMA, \S\ref{sss:hrma}), the objective transmission gratings (\S\ref{sss:gratings}), and the focal-plane science instruments (\S\ref{sss:fpsi}).

\subsubsection {High-Resolution Mirror Assembly (HRMA)} \label{sss:hrma}

\begin {figure} [htb]

\caption {\label{fig:hrma} 
Photograph of the High-Resolution Mirror Assembly (HRMA) during alignment and assembly in the HRMA Alignment Tower at Kodak.  In the picture, 7 of the 8 mirrors are already attached to the center aperture plate.  Photograph is from Kodak.}
\end {figure}

Hughes Danbury Optical Systems (HDOS, Danbury, Connecticut) - now Raytheon Optical Systems Incorporated (ROSI) - precision figured and superpolished the 4-mirror-pair grazing-incidence x-ray optics out of Zerodur blanks from Schott Glaswerke (Mainz, Germany).
Optical Coating Laboratory Incorporated (OCLI, Santa Rosa, California) coated the optics with iridium, chosen for high reflectivity and stability.
The Eastman Kodak Company (Rochester, New York) aligned and assembled the mirrors into the 10-m focal length High-Resolution Mirror Assembly (HRMA, Figure~\ref{fig:hrma}), which also includes thermal pre- and post-collimators and forward and aft contamination covers.
The forward contamination cover houses 16 radioactive sources, developed by MSFC, for verifying transfer of the flux scale from the ground to orbit~\cite{Elsner1994,Elsner1998,Elsner2000}.

\subsubsection {Objective transmission gratings} \label{sss:gratings}

Aft of the HRMA are 2 objective transmission gratings (OTGs) - the Low-Energy Transmission Grating (LETG) and the High-Energy Tranmission Grating (HETG).
Positioning mechanisms may insert either OTG into the converging beam to disperse the x-radiation onto the focal plane producing high-resolution spectra read-out by one of the focal-plane detectors (\S\ref{sss:fpsi}).

\paragraph {Low-Energy Transmission Grating (LETG)}

The Space Research Institute of the Netherlands (SRON, Utrecht, Netherlands) and the Max-Planck-Instit\"ut f\"ur extraterrestrische Physik (MPE, Garching, Germany) designed and fabricated the Low-Energy Transmission Grating (LETG, Figure~\ref{fig:gratings}).
The 540 grating facets, mounted 3 per module, lie tangent to the Rowland toroid  which includes the focal plane.
With free-standing gold bars of about 991-nm period, the LETG provides high-resolution spectroscopy from 0.08 to 2 keV (15 to 0.6 nm).

\begin {figure} [htb]

\caption {\label{fig:gratings} 
Photograph of the LETG and HETG mounted to the spacecraft structure.  Photograph is from TRW.}
\end {figure}

\paragraph {High-Energy Transmission Grating (HETG)}

The Massachusetts Institute of Technology (MIT, Cambridge, Massachusetts) designed and fabricated the High-Energy Transmission Grating (HETG, Figure \ref{fig:hetg}).
The HETG employs 2 types of grating facets~--- the Medium-Energy Gratings (MEG), mounted behind the HRMA's 2 outermost shells, and the High-Energy Gratings (HEG), mounted behind the HRMA's 2 innermost shells~--- oriented at slightly different dispersion directions.
With polyimide-supported gold bars of 400-nm and 200-nm periods, respectively, the HETG provides high-resolution spectroscopy from 0.4 to 4 keV (MEG, 3 to 0.3 nm) and from 0.8 to 8 keV (HEG, 1.5 to 0.15 nm).

\begin {figure} [htb]

\caption {\label{fig:hetg} 
Photograph of the High-Energy Transmission Grating (HETG).
Photograph is from the HETG team.}
\end {figure}

\subsubsection {Focal-plane science instruments} \label{sss:fpsi}

The ISIM (\S \ref{sss:sisys}) houses Chandra's 2 focal-plane science instruments~--- the (microchannel-plate) High-Resolution Camera (HRC) and the Advanced CCD Imaging Spectrometer (ACIS).
Each instrument provides both a socalled (as all the detectors are imagers) imaging detector (I) and a spectroscopy detector (S), the latter designed for reading out the high-resolution spectra dispersed by the Chandra's Observatory's insertable OTGs~--- HRC-S with the LETG and ACIS-S with the HETG. 

\paragraph {High-Resolution Camera (HRC)}

The Smithsonian Astrophysical Observatory (SAO, Cambridge, Massachusetts) designed and fabricated the High-Resolution Camera\cite{Murray1997} (HRC, Figure~\ref{fig:hrc}).
Made of a single large-format (10-cm-square) microchannel plate, the HRC-I provides high-resolution imaging over a large (31-arcmin-square) field of view.
Comprising 3 rectangular segments (3-cm-by-10-cm each) mounted end-to-end along the OTG dispersion direction, the HRC-S serves as the primary read-out detector for the LETG.
Both detectors are coated with a cesium--iodide photocathode and covered with aluminized-polyimide UV/ion shields.

\begin {figure} [htb]

\caption {\label{fig:hrc} 
Photograph of the focal plane of the Chandra flight High-Resolution Camera (HRC).
The HRC-I (imager) is at the bottom; the HRC-S (spectroscopic read-out), at the top.
Photograph is from the HRC team.}
\end {figure}

\paragraph {Advanced CCD Imaging Spectrometer (ACIS)}

The Pennsylvania State University (PSU, University Park, Pennsylvania) and the Massachusetts Institute of Technology (MIT, Cambridge, Massachusetts) designed and fabricated the ACIS (Figure~\ref{fig:acis}), with charge-coupled devices (CCDs) produced by MIT Lincoln Laboratory (Lexington, Massachusetts) and some subsystems and systems integration provided by Lockheed--Martin Astronautics (Littleton, Colorado).
Made of a 2-by-2 array of large-format (2.5-cm-square) CCDs, the ACIS-I provides high-resolution spectrometric imaging over a 17-arcmin-square field of view.
The ACIS-S, a 6-by-1 array of the large-format CCDs mounted along the OTG dispersion direction, serves both as the primary read-out detector for the HETG, and,  using BI CCD located at the center of the array, also provides high-resolution spectrometric imaging extending to lower energies but over a smaller (8-arcmin-square) field than ACIS-I.
Both ACIS detectors are covered with aluminized-polyimide optical blocking filters.

\begin {figure} [htb]

\caption {\label{fig:acis} 
Photograph of the focal plane of the Chandra flight Advanced CCD Imaging Spectrometer (ACIS), prior to installation of the optical blocking filters.
The ACIS-I (imager) is at the bottom; the ACIS-S (spectroscopic read-out), at the top.
Photograph is from the ACIS team.}
\end {figure}

\subsection {Ground system} \label{ss:ground}

The Chandra ground system comprises the Deep-Space Network (DSN, \S \ref{sss:dsn}), the Chandra Operations Control Center (OCC, \S \ref{sss:occ}), and the Chandra X-Ray Center (CXC, \S \ref{sss:cxc}).

\subsubsection {Deep-Space Network (DSN)} \label{sss:dsn}

The Jet Propulsion Laboratory (JPL, managed for NASA by the California Institute of Technology, Pasadena, California) operates NASA's Deep-Space Network (DSN).
Through its 3 antenna stations (Goldstone, California; Madrid, Spain; and Canberra, Australia), the DSN communicates directly with the Chandra spacecraft, up-linking commands and down-linking telemetered science and engineering data.
During normal operations, DSN contacts, at 8-hour intervals, the Chandra spacecraft, which stores 16.8 hours (1.8 Gb) of data in its solid-state recorder(s).

\subsubsection {Operations Control Center (OCC)} \label{sss:occ}

The Chandra Operations Control Center (OCC) is located in Cambridge, Massachusetts, and is part of the Chandra X-Ray Center (CXC, \S~\ref{sss:cxc}).
The Chandra prime contractor, TRW, staffs the OCC's Flight Operations Team (FOT), which is responsible for the control, health, and safety of the spacecraft.
The OCC receives Observation Requests, which it uses to build the Detailed Operations Timeline, and command loads to be sent to the spacecraft through the DSN.
From the spacecraft, through the DSN, the OCC receives the telemetered data, converts its format, extracts the engineering stream, and analyses the engineering data.
The OCC utilizes software developed by MSFC and by the Computer Sciences Corporation (CSC, El Segundo, California).

\subsubsection {Chandra X-Ray Center (CXC)} \label{sss:cxc}

The Smithsonian Astrophysical Observatory (SAO), with the Massachusetts Institute of Technology (MIT), operates the Chandra X-Ray Center (CXC).
As Chandra's day-to-day interface with the scientific community, the CXC supports NASA for soliciting observing proposals, provides proposal-preparation information and tools, organizes peer reviews on behalf of NASA, makes the long-term schedule from the approved observing proposals, generates Observation Requests (OR) from the long-term schedule and calibration needs, and submits the OR to the OCC for scheduling.
The CXC receives the re-formatted telemetry from the OCC, extracts the science stream, processes the science data, constructs time-tagged event lists for each observation, performs higher level data processing (\eg generates images and spectra), operates the data archive, maintains (with NASA Project Science) the calibration, and provides data, analysis tools, and other support to users.
The CXC is also responsible for developing and maintaining the software to support its functions.

\section {Ground calibration} \label{s:calibration}

The calibration of Chandra included an intensive and extensive ground calibration program for calibrating the full Observatory (\S \ref{ss:cal_grd}) and its individual subsystems.
The on-ground, and now on-orbit, calibration results (\S \ref{ss:cal_results}) clearly demonstrate that the Chandra Observatory provides the science capabilities (\S~\ref{s:onorbit})~--- high-resolution (sub-arcsec) imaging and spectrometric imaging and high-resolution dispersive spectroscopy~--- to address its science objectives.
With a goal of a high accurate calibration, the calibration program is an on-going effort which requires continued analysis and interpretation.
Our previous overviews~\cite{Weisskopf1997,O'Dell1998} provide additional details and references. 

\subsection {Observatory ground-calibration} \label{ss:cal_grd}

From 1996 December until 1997 May, the Chandra teams calibrated the Chandra Observatory at the MSFC X-Ray Calibration Facility (XRCF, Figure~\ref{fig:xrcf})~\cite{Kolodziejczak1995,Swartz1998}.
Calibration of the the Observatory used, of course, the flight High-Resolution Mirror Assembly (HRMA, \S~\ref{sss:hrma}) and flight objective transmission gratings (OTGs, \S~\ref{sss:gratings}), the Low Energy Transmission Grating (LETG) and the High Energy Transmission Grating (HETG), and the flight focal-plane detectors ACIS (\S~\ref{sss:fpsi}) and HRC (\S~\ref{sss:fpsi}).

\begin {figure} [htb]

\caption {\label{fig:xrcf} 
Aerial photograph of the X-Ray Calibration Facility (XRCF) at NASA Marshall Space Flight Center (MSFC).
The small building to the far left houses the X-ray Source System (XSS); the large building to the near right houses the instrument chamber, and control room, and provides space for data processing. 
MSFC photograph.}
\end {figure}

\vspace{0.10in}

\subsection {Summary of ground results} \label{ss:cal_results}

The Chandra science teams provided detail results of the ground calibration in various Calibration Reports, accessible, for example, through the Project Science calibration web pages (see~\ref{app:websites}). 
In addition to the Calibration Reports, some of the results appear in these~\cite{Jerius2000,Schwartz2000,Gaetz2000,Bautz2000,Murray2000,Kraft2000,Brinkman2000,Kenter2000,Pease2000,Prigozhin2000,Elsner2000} and previous~\cite{Zhao1998,Nousek1998,Chartas1998,Dewey1998,Schultz1998,VanSpeybroeck1997,Gaetz1997,Kenter1997,Kraft1997,Kellogg1997,Kolodziejczak1997,Brinkman1997,Predehl1997,Marshall1997,Dewey1997,Kolodziejczak1995,Swartz1998} proceedings.

Perhaps the most outstanding anomaly from the ground calibration had been a discrepancy (at about the 10-15\% level) between the measured HRMA effective area and the modelled (predicted) effective area at moderate (few keV) to high (several keV) energies~\cite{Schwartz2000}.
The discrepancy was speculated, and now confirmed, to have arisen from different and inconsistent approaches used in modelling the effects of surface roughness.
The synchrotron program to determine the optical constants of iridium-coated flats had used one technique, while the analysis of the Chandra HRMA calibration had used another.
The measured effective area (Figure~\ref{fig:area}) is now in reasonable agreement with model predictions.
Details are presented by Schwartz et al.~\cite{Schwartz2000} elsewhere in these proceedings.
\vspace{0.10in}

\begin {figure} [t]

\caption {\label{fig:area} 
Comparison of the model of the HRMA effective area with various data obtained during the Observatory calibration at the XRCF.  Plots are from Telescope Science and Mission Support Teams at SAO.}
\end {figure}

\section {On-orbit Performance} \label{s:onorbit}

Chandra's mission is to provide high-quality x-ray data.
In this section, we summarize certain key performance capabilities and address the degree to which they have been accomplished.

\subsection {Capabilities} \label{ss:capable}

Chandra is a unique x-ray astronomy facility for high-resolution imaging (\S~\ref{sss:imaging}) and for high-resolution spectroscopy (\S~\ref{sss:spec}).
Indeed, Chandra's performance advantage over other x-ray observatories is analagous to that of the Hubble Space Telescope (HST) over ground-bases observatories.

\subsubsection {Imaging performance} \label{sss:imaging}

The angular resolution of Chandra is significantly better than any previous, current, or even currently-planned x-ray observatory.
Figure~\ref{fig:casa} qualitatively, yet dramatically, illustrates this point by comparing the early Chandra image of the supernova remnant Cassiopeia-A, based on about 2700 s of data, with a $\sim$200,000 s ROSAT image.
(Prior to the development necessary to produce the Chandra optics, the ROSAT observatory represented the state of the art in high-resolution x-ray imaging.)
The improvement brought by Chandra's advance in angular resolution is dramatic, and the point source at the center~--- undetected in the ROSAT image~--- simply leaps out of the Chandra image.
\vspace{0.10in}

\begin {figure} [htb]
\vspace{0.10in}

\caption {\label{fig:casa} 
Chandra (left) and ROSAT (right) images of CAS-A.}
\end {figure}

Quantitatively, Chandra's point spread function (PSF), as measured during ground calibration, had a full width at half-maximum (FWHM) less than 0.5 arcsec and a half-power diameter (HPD) less than 1 arcsec.
The prediction for the on-orbit encircled-energy fraction was that a 1-arcsec-diameter circle would enclose at least half the flux from a point source.
The relatively mild dependence on energy (resulting from diffractive scattering by surface microroughness) attested to the excellent superpolished finish of the Chandra optics.
The ground measurements were, of course, taken under environmental conditions quite different than those encountered on-orbit.
Most notably the effects of gravity on the optics and the finite distance and size of the various x-ray sources used were unique to the ground calibration.
On the other hand, on the ground there was no Observatory motion to deal with.
On-orbit the performance folds in the spatial resolution of the flight detectors and any uncertainties in the aspect solution which determines, post-facto, the direction the observatory was pointing relative to the instruments and to celestial coordinates.

The High Resolution Camera (HRC) has the best spatial resolution  ($\sim$ 20$\mu$m, $\sim$0.4 arcsec) of the two imaging instruments aboard Chandra and thus is best matched to the telescope.
Figure~\ref{fig:eehrci} illustrates the extrapolation of the ground calibration to on-orbit performance and compares the predictions at two energies with an observed PSF.
Figure~\ref{fig:eehrcs} shows a similar comparison using the HRC-S.
The angular resolution of the Chandra X-Ray Observatory has been as expected.
Further details concerning the Chandra point-spread function are presented by Jerius and colleagues in these proceedings~\cite{Jerius2000}.
The on-orbit performance of the HRC is discussed in more detail by Murray et al.~\cite{Murray2000}, Kenter et al.~\cite{Kenter2000}, and Kraft et al.~\cite{Kraft2000}.
Similarly, these proceedings contain more detailed discussion of the performance of the aspect camera and the attitude control in the papers by Aldcroft et al.~\cite{Aldcroft2000} and Cameron et al.~\cite{Cameron2000}.

\begin {figure} [htb]

\caption {\label{fig:eehrci} 
The predicted and observed encircled energy as a function of radius for an on-axis point source.
The detector is the HRC-I.
The calculations, performed at two energies~--- 0.277 keV and 6.40 keV, include a realistic (0.22") estimate of the contribution from the aspect solution.
Flight data from the calibration observation of AR Lac are also shown.
Figure produced by Telescope Science.}
\end {figure}

\begin {figure} [htb]

\caption {\label{fig:eehrcs} 
The predicted and observed encircled energy as a function of radius for an on-axis point source.
The detector is the HRC-S.
The calculations, performed at two energies~--- 0.277 keV and 6.40 keV, include a realistic (0.22") estimate of the contribution from the aspect solution.
Flight data from the calibration observation of SMC X-1 are also shown.
Figure produced by Telescope Science.}
\end {figure}

\subsubsection {Spectroscopic performance} \label{sss:spec}

The unprecedented angular resolution of the Chandra optics, combined with Chandra's micro-ruled objective transmission gratings (OTGs), provides the capability for high-resolution dispersive spectroscopy.
Chandra has two sets of OTGs~--- the Low-Energy Tranmission Grating (LETG) is optimized for longer x-ray wavelengths, and the High-Energy Tranmission Grating (HETG) for shorter wavelengths.
Hence, with an appropriate combination of Chandra's gratings, Chandra allows measurements with spectral resolving power (Figure~\ref{fig:gres}) of $(\lambda/\Delta\lambda = (E/\Delta~E) > 500$ for wavelengths $\lambda > 0.4$ nm (energies $<$ 3 keV).

\begin {figure} [htb]

\caption {\label{fig:gres}  
Specified spectral resolving power of Chandra OTGs.
Preliminary results indicate slightly better performance.
Plot is from the Chandra Project Science team.}
\end {figure}

\subsection {Performance anomalies} \label{ss:anomalies}

The performance of the Chandra Observatory, the instruments and subsystems have been remarkably free from problems and anomalies.
Here we discuss the two difficulties that have been encountered that have had some impact on the scientific performance.
We note, however, that {\sl neither are preventing the mission from accomplishing its scientific objectives}.

\subsubsection {Proton damage to the front-illuminated CCDs} \label{damage}

The ACIS front-illuminated CCDs originally approached the theoretical limit for the energy resolution at almost all energies, while the back-illuminated devices exhibited poorer resolution.
Subsequent to launch and orbital activation, the energy resolution of the front-illuminated (FI) CCDs has become a function of the row number, being nearer pre-launch values close to the frame store region and progressively degraded towards the farthest row.
An illustration of the current dependence on row is shown in Figure~\ref{fig:rows}.

\begin {figure} [htb]

\caption {\label{fig:rows}  
The energy resolution of S3 and I3 as a function of row number.
These data were taken at -120$^o$C.
Note that these curves are representative of the variation~--- but they do not account for the row-dependent gain variation which also increases the energy resolution by an additional 15-20\% for the larger row numbers.}
\end {figure}

For a number of reasons, we believe that the damage was caused by low energy protons, encountered during radiation belt passages and reflecting off the x-ray telescope onto the focal plane.
Subsequent to the discovery of the degradation, operational procedures were changed and the ACIS is not left at the focal position during radiation belt passages.
(The HRC is left at the focal position, but with its door partially closed for protection.)
Since this procedure was initiated, no further degradation in performance has been encountered.
The BI CCDs were not impacted and this result is consistent with the proton-damage scenario as it is far more difficult for low energy protons from the direction of the HRMA to deposit their energy in the buried channels (where damage is most detrimental to performance) of the BI devices, since these channels are near the gates and the gates face in the direction opposite to the HRMA.
Thus the energy resolution for the two BI devices remains at their prelaunch values.

The position dependent energy resolution of the FI chips depends on the ACIS operating temperature.
Since activation, the ACIS operating temperature  has been slightly lowered, based on considerations of molecular contamination, and is now set at the lowest temperature now thought safely achievable ($\sim -120^o$C).

More recently, the ACIS team has been able to reproduce the damage characteristics after bombarding test devices with low-energy protons ($<$ few hundred keV).
Furthermore, by sweeping charge through the system they have beeen able to fill the charge traps and further dramatically reduce the impact.
On-orbit testing of this technique will take place in the near future (April 2000).
Further details are provided elsewhere in these proceedings~\cite{Bautz2000,Prigozhin2000}.

\subsubsection {HRC-S Anticoincidence Electronics} \label{sss:anit}

The anti-coincidence shield of the HRC-S is not working because of a timing error in the electronics.  
The error is not correctable.
As a result the raw event rate is very high and exceeds the total telemetry rate limit.
To cope with this the HRC Team has defined a "spectroscopy region" which is about 1/2 of the width and extends along the full length of the HRC-S detector.
With this change, the quiescent on-orbit background rate is about 85 cts s$^{-1}$.
This background can be further reduced in ground data processing by using pulse height filtering that preferentially selects x-rays over the cosmic ray events.
A further reduction in background of a factor of about three is possible.
More details on the performance of the HRC may be found elsewhere in these proceedings~\cite{Murray2000,Kenter2000,Kraft2000}.

\subsection {Scientific Performance} \label{ss:sciperf}

X-rays result from highly energetic processes - thermal processes in plasmas with temperatures of millions of degrees or nonthermal processes, such as synchrotron emission or scattering from very hot or relativistic electrons.
Consequently, x-ray sources are frequently exotic:

\begin{itemize}
\item Supernova explosions and remnants, where the explosion shocks the ambient interstellar medium or a pulsar (rotating neutron star) powers the emission.
\item Accretion disks or jets around stellar-mass neutron stars or black holes.
\item Accretion disks or jets around massive black holes in galactic nuclei.
\item Hot gas in clusters of galaxies and in galaxies, which traces the gravitational field for determining the mass.
\item Hot gas in stellar coronae, especially during flares (coronal mass ejection).
\end{itemize}

Here we give a few examples of observations with Chandra which indicate the potential for investigating these processes and astronomical objects through high-resolution imaging (\S~\ref{sss:sciimaging}) and high-resolution spectroscopy (\S~\ref{sss:scispec}).

\subsubsection {Imaging} \label{sss:sciimaging}

Chandra's capability for high-resolution imaging (\S~\ref{sss:imaging}) enables detailed high-resolution studies of the structure of extended x-ray sources, including supernova remnants (Figure~\ref{fig:casa}), astrophysical jets (Figure~\ref{fig:pks0637}, and hot gas in galaxies and clusters of galaxies (Figure~\ref{fig:hydra}).
The supplementary capability for spectrometric imaging allows studies of structure, not only in x-ray intensity, but in temperature and in chemical composition.
Through these observations, astronomers will address several of the most exciting topics in contemporary astrophysics~--- \eg galaxy mergers, dark matter, and the cosmological distance scale.

\begin {figure} [htb]

\caption {\label{fig:pks0637} 
X-ray image of the source PKS0637 with radio contours overlaid.
The distance from the central object to the x-ray bright knot is $\sim$10".
Image courtesy CXC.}
\end {figure}

\begin {figure} [htb]

\caption {\label{fig:hydra} 
X-ray image of the center of the Hydra cluster of galaxies.
Image courtesy CXC.}
\end {figure}

In addition to mapping the structure of extended sources, the high angular resolution permits studies of ensembles of discrete sources, which would otherwise be impossible owing to source confusion.
A beautiful example comes from the recent observations of the center of M31 (Figure~\ref{fig:m31}) performed by M. Garcia and colleagues~\cite{Garcia2000}.
The image shows what used to be considered as emission associated with the black hole at the center of the galaxy now resolved into several distinct objects.
A most interesting consequence is that the emission from the region surrounding the central black hole is now known to be dramatically reduced and unexpectedly and surprisingly faint!
Thus, Chandra observations will isolate individual stars in clusters and star-forming regions and x-ray binaries in nearby normal galaxies.
Furthermore, high-angular-resolution observations with Chandra's low-noise focal-plane detectors will obtain photon-limited, deep-field exposures which are likely to resolve most of the extragalactic cosmic x-ray background into faint, discrete sources~\cite{Mushotzky2000} (Figure~\ref{fig:xrb}).
\vspace{0.10in}

\begin {figure} [htb]

\caption {\label{fig:m31} 
X-ray image of the center of the galaxy M31.
Image courtesy S. Murray.}
\end {figure}

\begin {figure} [htb]

\caption {\label{fig:xrb} 
A moderately deep ($\sim$100 ksec) Chandra image showing dozens of faint x-ray sources.
Image courtesy R. Mushotzky.}
\end {figure}

\subsubsection {Spectroscopy} \label{sss:scispec}

Owing to their unprecedented clarity, Chandra images will be visually striking and provide new insights into the nature of x-ray sources.
Equally important to the imaging science (\S~\ref{sss:sciimaging}) will be Chandra's unique contributions to high-resolution dispersive spectroscopy.
Indeed, as the capability for visible-light spectroscopy begat the field of astrophysics about a century ago, high-resolution x-ray spectroscopy will contribute profoundly to the understanding of the physical processes in cosmic x-ray sources.

High-resolution x-ray spectroscopy is the essential tool for diagnosing conditions in hot plasmas.  It provides information for determining the temperature, density, elemental abundance, and ionization stage of x-ray emitting plasma.
The high spectral resolution of Chandra isolates individual lines from the myriad of spectra lines which would overlap at lower resolution.
Furthermore, it enables the determination of flow and turbulent velocities, through measurement of Doppler shifts and widths.

Dispersive spectroscopy achieves its highest resolution for spatially unresolved (point) sources.
Thus, OTG observations will concentrate on stellar coronae, x-ray binaries, and active galactic nuclei.
Figure~\ref{fig:hr1099a} illustrates with an ACIS-S image of the spectra dispersed by the HETG during observations of HR1099.  
Figure~\ref{fig:hr1099b} shows the line rich extracted spectrum.
Figure~\ref{fig:capella} shows a similarly extracted spectrum from LETG observations of Capella.

\begin {figure} [htb]

\caption {\label{fig:hr1099a} 
Image of the spectra dispersed by the HETG during observations of HR1099.
Image courtesy Dan Dewey and HETGS team.}
\end {figure}

\begin {figure} [htb]

\caption {\label{fig:hr1099b} 
The MEG spectrum of HR1099.
Image courtesy Dan Dewey and HETGS team.}
\end {figure}

\begin {figure} [htb]

\caption {\label{fig:capella} 
Capella spectrum from LETGS.
Image courtesy Jeremey Drake and LETGS team.}
\end {figure}

It is interesting that the use of the zeroth order image for the observations of extremely bright sources, which would otherwise saturate the detectors and/or the telemetry, has proven quite useful.
The utility for such observations is illustrated in Figure~\ref{fig:3c273} where both the jet and the central source of 3C273 are clearly resolved.

\begin {figure} [htb]

\caption {\label{fig:3c273} 
Image of the dispersed spectrum, including zeroth order, of 3C273.
The jet is clearly resolved in the lower right hand portion of the figure.
The six spikes emanating from the central image are due to dispersion by the facet holders.
Image courtesy Jeremey Drake and LETGS team.}
\end {figure}

Besides observing the emission spectra of cosmic sources, Chandra high-resolution spectroscopy will also probe the interstellar medium through its absorption of x-rays from bright sources.
Such observations of absorption spectra provide information on both interstellar gas and dust~--- the latter through the analysis of extended x-ray absorption fine sturcture~\cite{Evans1986,Woo1995} (EXAFS) and x-ray absoprtion near-edge structure~\cite{Woo1995} (XANES).

\section {Conclusion} \label{s:conc}

The Chandra X-Ray Observatory is performing as well, if not better, than anticipated~--- and the results are serving to usher in a new age of astronomical and astrophysical discoveries.

\section {Acknowledgements} \label{s:ackn}

We recognize the efforts of the various Chandra teams which have contributed to the success of the observatory.
In preparing this overview, we have used figures drawn from their work.

\appendix
\section{Chandra web sites} \label{app:websites}
The following lists several Chandra-related sites on the World-Wide Web (WWW). Most sites are cross-linked to one another.

\begin{description}

\item[\verb|http://chandra.harvard.edu/|] Chandra X-Ray Center (CXC), operated for NASA by the Smithsonian Astrophysical Observatory (SAO).

\item[\verb|http://wwwastro.msfc.nasa.gov/xray/axafps.html|] Chandra Project Science, at the NASA Marshall Space Flight Center (MSFC).

\item[\verb|http://hea-www.harvard.edu/HRC/|] Chandra High-Resolution Camera (HRC) team, at the Smithsonian Astrophysical Observatory (SAO).

\item[\verb|http://www.astro.psu.edu/xray/axaf/axaf.html|] Advanced CCD Imaging Spectrometer (ACIS) team at the Pennsylvania State University (PSU).

\item[\verb|http://acis.mit.edu/|] Advanced CCD Imaging Spectrometer (ACIS) team at the Massachusetts Institute of Technology (MIT).

\item[\verb|http://www.ROSAT.mpe-garching.mpg.de/axaf/|] Chandra Low-Energy Transmission Grating (LETG) team at the Max-Planck Instit\"ut f\"ur extraterrestrische Physik (MPE).

\item[\verb|http://space.mit.edu/HETG/|] Chandra High-Energy Transmission Grating (HETG) team, at the Massachusetts Institute of Technology (MIT).

\item[\verb|http://hea-www.harvard.edu/MST/|] Chandra Mission Support Team (MST), at the Smithsonian Astrophysical Observatory (SAO).

\item[\verb|http://ipa.harvard.edu/|] Chandra Operations Control Center, operated for NASA by the Smithsonian Astrophysical Observatory (SAO).

\item[\verb|http://ifkki.kernphysik.uni-kiel.de/soho|] EPHIN particle detector.

\end{description}

\bibliography{/axaf/papers/axaf}
\bibliographystyle{spiebib}

\end{document}